\documentclass[sigconf]{acmart}
\usepackage{soul}
\usepackage{enumitem}
\usepackage{subcaption}
\usepackage{multirow}

\AtBeginDocument{%
  \providecommand\BibTeX{{%
    \normalfont B\kern-0.5em{\scshape i\kern-0.25em b}\kern-0.8em\TeX}}}

\copyrightyear{2025}
\acmYear{2025}
\setcopyright{cc}
\setcctype{by}
\acmConference[AISec '25]{Proceedings of the 2025 Workshop on Artificial Intelligence and Security}{October 13--17, 2025}{Taipei, Taiwan}
\acmBooktitle{Proceedings of the 2025 Workshop on Artificial Intelligence and Security (AISec '25), October 13--17, 2025, Taipei, Taiwan}\acmDOI{10.1145/3733799.3762977}
\acmISBN{979-8-4007-1895-3/2025/10}

\begin{document}


\title{Ensembling Membership Inference Attacks Against Tabular Generative Models}


\author{Joshua Ward}
\email{joshuaward@ucla.edu}
\affiliation{
 \institution{University of California Los Angeles}
  \city{Los Angeles}
  \state{California} 
 \country{USA}
}
\author{Yuxuan Yang}
\email{christyyxyang@gmail.com }
\affiliation{
 \institution{Stanford University}
   \city{Palo Alto}
  \state{California} 
 \country{USA}
}
\author{Chi-Hua Wang}
\email{chihuawang@ucla.edu}
\affiliation{
 \institution{University of California Los Angeles}
   \city{Los Angeles}
  \state{California} 
 \country{USA}
}

\author{Guang Cheng}
\email{guangcheng@ucla.edu}
\affiliation{
 \institution{University of California Los Angeles}
   \city{Los Angeles}
  \state{California} 
 \country{USA}
}



\renewcommand{\shortauthors}{Ward, et al.}

\begin{abstract}

Membership Inference Attacks (MIAs) have emerged as a principled framework for auditing the privacy of synthetic data generated by tabular generative models, where many diverse methods have been proposed that each exploit different privacy leakage signals. However, in realistic threat scenarios, an adversary must choose a single method without a priori guarantee that it will be the empirically highest performing option. We study this challenge as a decision theoretic problem under uncertainty and conduct the largest synthetic data privacy benchmark to date. Here, we find that no MIA constitutes a strictly dominant strategy across a wide variety of model architectures and dataset domains under our threat model. Motivated by these findings, we propose ensemble MIAs and show that unsupervised ensembles built on individual attacks offer empirically more robust, regret-minimizing strategies than individual attacks. \footnote{A code repository can be found at: \href{https://github.com/joshward96/Ensemble-MIA}{github.com/joshward96/Ensemble-MIA.} }

\end{abstract}





\keywords{Tabular Synthetic Data, Membership Inference Attack, Privacy}



\maketitle

\section{Introduction}
Tabular data synthesis has emerged as a methodology that has demonstrated success in private data release \cite{privbayes,yoon2018pategan,yoon2020anonymization}, training dataset augmentation for supervised learning \cite{cui2024tabulardataaugmentationmachine}, and missing value imputation \cite{zheng2023diffusionmodelsmissingvalue, liu2024self}. As organizations increasingly rely on synthetic data to balance utility with privacy concerns, the ability to generate high-quality tabular datasets that preserve statistical properties while protecting individual privacy has become critical. However, many popular tabular generative model implementations including Generative Adversarial Networks \citep{Xu2019ModelingTD,yoon2018pategan,yoon2020anonymization}, language models \citep{borisov2023languagemodelsrealistictabular, solatorio2023realtabformer}, and Diffusion models \citep{tabddpm,autodiff,tabsyn}, do not provide formal privacy guarantees despite their widespread adoption. While these methods can generate synthetic data that maintains distributional characteristics of original datasets, the privacy protection they offer is largely implicit or argued through non-adversarial methodologies such as similarity metrics.

Membership Inference Attacks (MIAs) are a primary methodology for auditing the privacy of tabular generative models that attempt to determine whether a specific record was part of the training dataset. These attacks serve as a practical tool for evaluating privacy leakage, as they present privacy auditing as a game where an adversary, given a threat model that describes what information can be used, constructs an attack that classifies whether a test observation is a member of the dataset a model was trained with. A successful attack represents a practical and interpretable privacy breach. As a classic example, an insurance company could have access to a hospital's synthetic cancer dataset and, for a new applicant, attack the dataset to determine if the applicant is a member, leaking their diagnosis \citep{hu2022membershipinferenceattacksmachine}. MIAs have often been used for privacy assessment \cite{kazmi2024panoramiaprivacyauditingmachine, mireshghallah2022quantifyingprivacyrisksmasked} and differentially private algorithm auditing \cite{Jagielski2020, annamalai2025hitchhikersguideefficientendtoend}.

While the general MIA methodology is well-established, specific attacks for tabular synthetic data vary considerably in their approach, targeting different privacy leakage signals and exploiting distinct aspects of model memorization. For example, some attacks focus on leveraging statistical overfitting patterns while others evaluate evidence of memorization using nearest neighbor-based calculations. A danger is that different attacks may \textit{ underestimate} the actual privacy leakage of synthetic data or only perform well in different domains or under different generative model architectures. Additionally, if there is \textit{disagreement} among attacks, the individual privacy for a member becomes conditioned on whichever attack strategy an adversary chose to use.
Due to the high-leverage use cases for synthetic data in fields such as healthcare \cite{Vallevik_2024}, finance \cite{potluru2024syntheticdataapplicationsfinance}, and education \cite{liu2024scalingprivacypreservingcomprehensive}, which regularly use sensitive personal identification information, accurate and comprehensive privacy evaluation is critical for the deployment of trustworthy generative AI systems.

Motivated by this diversity in attack strategies, we frame the privacy auditing problem as a strategy selection challenge under an unknown state. Adversaries typically do not know which generative model and dataset combination they will encounter when deployed against a responsible defender. Since they can only select one strategy, the key question becomes: which attack strategy minimizes regret—that is, performs consistently well across different target model and dataset combinations? This leads to our first research question:
\begin{itemize}
\item \textbf{Research Question 1: Does there currently exist an MIA for tabular synthetic data generators that is a strictly dominant strategy across different generative models and datasets?}
\end{itemize}
In the largest tabular synthetic data privacy benchmark to date, we show that no single attack consistently outperforms others, indicating the absence of a strictly dominant strategy across synthetic data from 9 generative models and 57 datasets (see Figure \ref{fig:att_prop}). This variability makes it difficult for practitioners to select appropriate privacy auditing methods and suggests that relying on any single attack may provide an incomplete assessment of privacy risks. Indeed, we find that many attacks' scores are only weakly correlated with each other (see Figure \ref{fig:corr_plot}) and that attack disagreement is often significant.

The diversity of attack performance motivates us to explore methodologies that can provide more robust privacy auditing in the absence of a dominant strategy. Drawing inspiration from ensemble learning, where combining multiple weak learners often yields superior performance, we investigate whether treating individual membership inference attacks as components in an ensemble can create better regret-minimizing strategies. The intuition is that different attacks may capture complementary privacy leakage signals, with each potentially excelling under different generative model architectures or data characteristics. This leads to our second research question:
\begin{itemize}
\item \textbf{Research Question 2: Can ensemble methods create more robust MIA strategies that minimize regret compared to individual attacks?}
\end{itemize}

Here, we show that ensembling individual MIAs consistently improves performance from a regret-minimizing standpoint, seeing better mean ranks over the benchmark relative to individual attacks (Table \ref{tab:performance_comparison}). This indicates that while ensembles are also neither strictly dominant, they are a more robust strategy for a rational adversary. We also show that individual attacks that do not have the best individual attack performance can contribute more to the success of an ensemble than their corresponding best individual strategy counterparts. These findings not only allow for better performing MIA strategies across broader tabular synthetic data domains, but they also highlight a promising research direction for MIAs even if an individual attack is not highest performing relative to its peers, if it is sufficiently uncorrelated it can be used to improve ensemble attacks.

\section{Background and Preliminaries}
\subsection{Tabular Synthetic Data Generation}
We denote tabular data as a matrix $\mathbf{X} \in \mathcal{X}^{n \times d}$, where $n$ represents the number of samples, $d$ the number of features, and $\mathcal{X}$ is the domain of possible feature values. Each row $\mathbf{x}_i \in \mathcal{X}^d$ corresponds to a single data point sampled from the underlying distribution $p_X(X)$, and each column represents a feature with potentially different data types. We use $\mathbf{x}_{i,j}$ to denote the value of the $j$-th feature for the $i$-th sample. A training dataset $T = {\mathbf{x}_1, \mathbf{x}_2, \ldots, \mathbf{x}_n}$ consists of $n$ independent samples drawn from $p_X(X)$.

The goal of tabular generative models is to learn a generative model $G$ from the training dataset $T$ that approximates the underlying data distribution $p_X(X)$. The model $G$ can then generate new synthetic samples $\tilde{\mathbf{x}} \sim G$ that form a synthetic dataset $S = {\tilde{\mathbf{x}}_1, \tilde{\mathbf{x}}_2, \ldots, \tilde{\mathbf{x}}_m}$. The synthetic data should preserve both marginal distributions of individual features and the complex joint dependencies between features present in the original distribution.

Unlike images or text, tabular data exhibits several unique properties that pose privacy challenges for generative modeling. First, tabular datasets typically contain heterogeneous feature types, including continuous numerical values, discrete categorical variables, and ordinal features. Second, the dimensionality is generally moderate (tens to hundreds of features) compared to other domains, but the relationships between features can be highly non-linear and complex. Third, tabular data often exhibits irregular distributions with skewness, multi-modality, and varying scales across features. Failure to model these characteristics well can lead to privacy leakage signals that MIAs can exploit.

\subsection{Membership Inference Attacks on Synthetic Data Generators}

Membership Inference Attacks (MIAs) aim to classify whether a specific observation was a member of the original dataset used to train a model. Given the generative model $G$ trained on dataset $T$ as defined above, which generates synthetic dataset $S$, an adversary $\mathcal{A}: X \to \{0, 1\}$ aims to determine if a test sample $x^*$ is an element of $T$. Formally, this classification or Membership Inference Attack can be expressed as:
\begin{equation}\label{eq:membership_prediction}
    \mathcal{A}(x^{\star}) = \mathbb{I}\left[f(x^{\star}) > \gamma\right]
\end{equation}
where $\mathbb{I}$ is the indicator function, $f(x^{\star})$ is a scoring function of the test observation $x^*$, and $\gamma$ is an adjustable decision threshold. The success of the attack can be measured using traditional binary classification metrics and can be interpreted as a measure of privacy leakage from a model of the training data.

 To construct their attack, the adversary relies on some prior information called a threat model. These include black box attacks \cite{Hayes2017LOGANMI, Hilprecht2019MonteCA, ganleaks} in which only $S$ is available,
  shadow box (also called calibrated) attacks in which both $S$ and then a reference dataset $R$ from the same population distribution of the training set are given \cite{vanbreugel2023membership,ward2024dataplagiarismindexcharacterizing, Gen-LRA}, and white box attacks \citep{sablayrolles2019white} in which both $S$, $R$ and full access to the model are known. Other lines of work have explored threat models where the adversary assumes a shadow-box threat model but additionally knows the implementation, but not the training weights, of the tabular generator \cite{groundhog, houssiau2022tapas,Meeus_2024}.

 MIAs leverage information from a specified threat model along with some hypothesis about model failure modes such as memorization or overfitting to exploit potential vulnerabilities in constructing Equation \ref{eq:membership_prediction}. 
 For example, a variety of attacks from \cite{ganleaks} and  \cite{houssiau2022tapas} target memorization by computing the distance between $x^*$ and the closest observation from $S$. Other MIAs focus on overfitting, where the model produces synthetic samples that are too similar in distribution to the training dataset relative to the overall population distribution. Methods such as DOMIAS \cite{vanbreugel2023membership}, DPI \cite{ward2024dataplagiarismindexcharacterizing} and Gen-LRA \cite{Gen-LRA}
 attack overfitting by comparing the density of synthetic observations in a local region to that of a reference dataset.

While methodologically diverse, MIAs targeting synthetic data aim to uncover the same fundamental issue: the potential for generative models to inadvertently reveal information about their training data. If a model produces synthetic records that allow an adversary to infer training membership, it constitutes a direct breach of privacy. This leakage signals a failure in the model, as it indicates an imbalance between generating realistic data and preserving confidentiality. A well-calibrated generative model should neither reproduce training samples nor generate synthetic data that is overly concentrated around specific regions of the training distribution.
 

\subsection{Threat Model} 
In this work, we specifically focus on "No-box" \cite{houssiau2022tapas} attacks, where the generator is assumed to be unknown and inaccessible and the adversary only has access to the released synthetic data $S$ and a reference dataset $R$ sampled from the same population distribution as $T$. These categories of attacks are particularly relevant as they target the privacy leakage inherent in the released synthetic dataset
  itself. We argue that these threat models should be the primary focus in the tabular data synthesis domain for the following reasons:
  
  \textbf{Plausibility}: No-box threat models are most proximate to the synthetic data release paradigm in which a practitioner wishes to release their synthetic data to the public or a selected group. In these circumstances, an adversary would only have access to this synthetic data and perhaps a reference dataset which could be obtained through domain knowledge, open-source information, or paid collection. \cite{vanbreugel2023membership} for example, showed that even artificial reference datasets constructed from histograms of population data can increase attack performance. This stands in contrast to "Model Known" black-box and shadow-box attacks, which are unsuitable for realistic threat modeling. These attacks are trivially easy to defeat, as the defender can simply choose not to release the implementation details of the generative model with $S$. Indeed, \cite{golob2024privacyvulnerabilitiesmarginalsbasedsynthetic} has shown that significant privacy leakage can occur in differentially private synthetic data generation when even the model implementation is disclosed. Therefore, the best practice for data-releasing parties is to disclose as little model information as possible, making model-agnostic No-box attacks the most relevant.

\textbf{Compatibility:} A key advantage of these threat models is that the corresponding attacks are definitionally compatible with all tabular generators, as they only assess the output of these models. This allows for fair benchmarking between both attacks and models and represents a data-centric approach like the corresponding utility metrics used for tabular data synthesis. Indeed if there exists an attack that only works for diffusion or language models, a savvy defender would just choose to not use those architectures.

\begin{table}[t]
\centering
\caption{Membership-inference attacks used in this study.}
\label{tab:mia_list}
\begin{tabular}{ll}
\toprule
\textbf{Attack} &  \textbf{Signal type} \\
\midrule
DOMIAS \cite{vanbreugel2023membership}           & Density ratio \\
DPI \cite{ward2024dataplagiarismindexcharacterizing}              & Local density \\
Classifier \cite{houssiau2022tapas}       & Density ratio \\
Gen-LRA \cite{Gen-LRA}       & Likelihood ratio \\

DCR \cite{ganleaks}                      & Distance-based \\
DCR-Diff \cite{ganleaks}                 & Distance difference \\
Logan \cite{Hayes2017LOGANMI}            & Density ratio \\
MC Estimation \cite{Hilprecht2019MonteCA}  & Density estimation \\
\bottomrule
\end{tabular}

\end{table}
\subsection{Considered Attacks}
Under this threat model, a wide variety of attacks have been proposed to audit the privacy of synthetic data that rely on different attack signals. We will use and reference these attacks throughout our paper.

 \textbf{Distance to Closest Record (DCR/ DCR-Diff)} Distance-based membership inference attacks \cite{ganleaks} operate on the hypothesis that synthetic data generators exhibit memorization behavior toward training data, resulting in synthetic records that are geometrically closer to member records than to non-member records in the feature space. The Distance to Closest Record (DCR) attack \cite{ganleaks} targets this by constructing Equation \ref{eq:membership_prediction} as: $f_{\text{DCR}}(x^*) = -\min_{\mathbf{x} \in S} d(x^*,\textbf{x})$ where $d(\cdot,\cdot)$ is some measure of distance. DCR-Diff builds on this idea by calibrating the attack with a holdout reference dataset that subtracts the distance of the nearest reference record: $f_{\text{DCR}}(x^*) = -\min_{\mathbf{x} \in S} d(x^*,\textbf{x})-\min_{\mathbf{x} \in R} d(x^*,\textbf{x})$.
 
\begin{figure*}
    \centering
    \includegraphics[width=.6\linewidth]{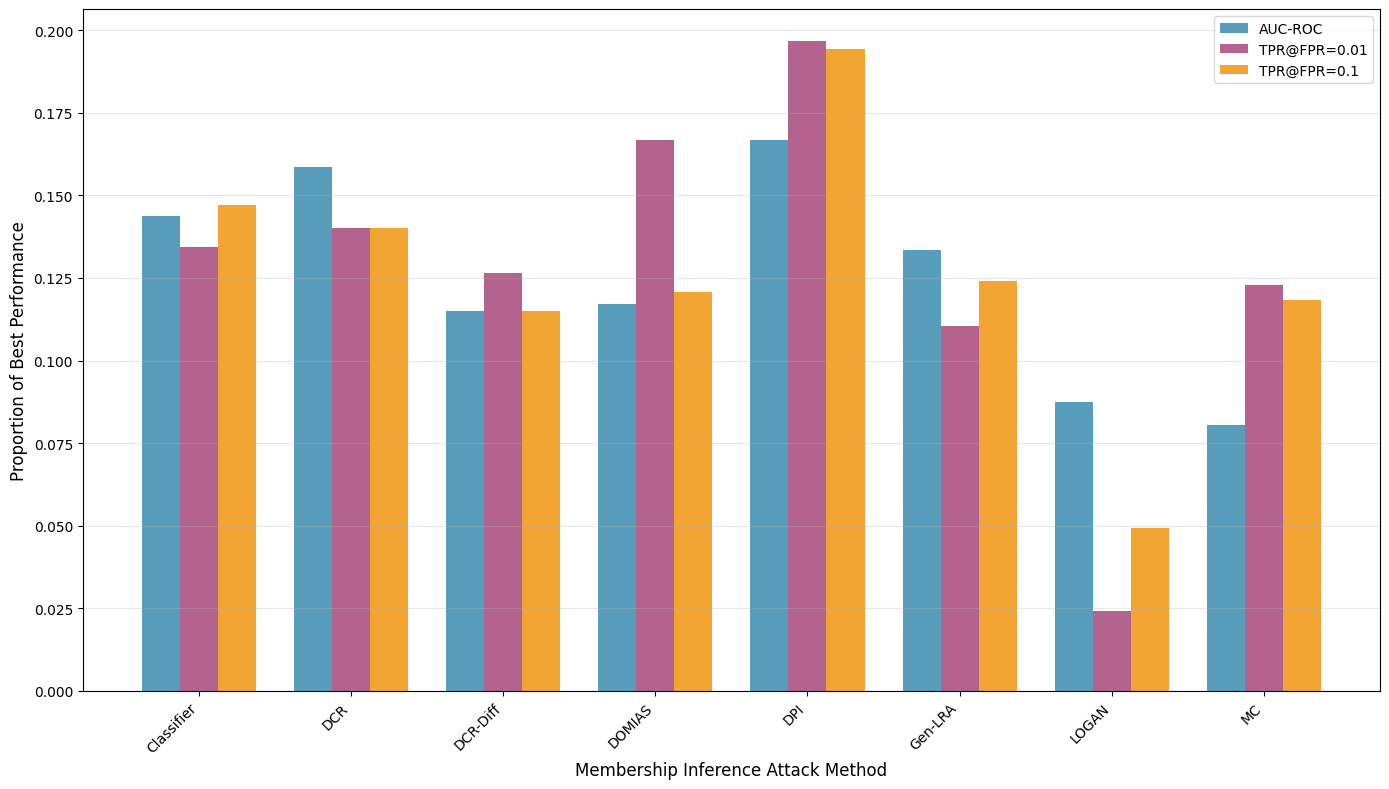}
    \caption{Proportion of instances each MIA had the highest AUC of all other attacks across all generative models, datasets, and seeds. The highest performing attack DPI is only the most successful in terms of AUC and TPR@FPR=0.1 in 16.2\% and 19.1\% of experiment runs respectively. This suggests that there is not a strictly dominant adversarial strategy across attacks with comparable threat models.}
    \label{fig:att_prop}
    \Description{A horizontal bar chart showing the proportion of experimental runs where each membership inference attack (MIA) achieved the highest AUC score. The bars represent different MIA methods, with DPI showing the highest proportion at approximately 16.2\%, followed by other methods with varying but generally lower proportions, demonstrating the relative performance distribution across different attack strategies.}

\end{figure*}
\textbf{DOMIAS.} The DOMIAS \cite{vanbreugel2023membership} attack employs a density-based methodology that finds signal by attacking model overfitting in the synthetic dataset. Here, DOMIAS computes the density ratio of $x^*$ over the estimated probability density functions of $S$ and $R$, creating a calibrated scoring function: $f_{\text{DOMIAS}}(x^*) = \frac{p_S(x^*)}{p_R{(x^*)}}$. DOMIAS requires estimating these densities separately and uses either Kernel Density Estimators or deep learning-based methods. 

\textbf{Data Plagiarism Index (DPI).} The Data Plagiarism Index attack \cite{ward2024dataplagiarismindexcharacterizing} quantifies local memorization behavior by analyzing the density ratio of synthetic versus reference data points in local neighborhoods. For each query record $x^*$, DPI constructs a K-nearest neighborhood $D(x^*)$ using both reference and synthetic data points, then computes the scoring function as the ratio of synthetic to reference points within this neighborhood: $f_{\text{DPI}}(x^*)  = \frac{\sum_{\mathbf{z} \in D(x^*)} \mathbb{I}(\mathbf{z} \in S)}{\sum_{\mathbf{z} \in D(x^*)} \mathbb{I}(\mathbf{z} \in R)}$. The DPI value provides interpretable results: DPI = 0 indicates under-fitting, DPI = 1 represents balanced generation, and DPI > 1 suggests memorization through disproportionate synthetic concentration.

\textbf{Gen-LRA.} Gen-LRA \cite{Gen-LRA} treats membership inference as evaluating the influence of $x^*$ on the likelihood of $S$ evaluated by a surrogate density estimator on $R$. The idea is that if the likelihood of $S$ is substantially higher under a model fit with the inclusion of $x^*$, there is evidence of overfitting. Gen-LRA further improves their attack by localizing the evaluation of $S$ to samples that are close in distance to $x^*$. The technique utilizes Gaussian Kernel Density Estimation (KDE) to approximate the required probability distributions, computing a likelihood ratio as the scoring function: $f_{\text{Gen-LRA}}(x^*) = \frac{\prod_{s \in S} {p}_{{R}\cup{x^*}}(s)}{\prod_{s \in S}{p}_{R}(s)}$.

\textbf{LOGAN/ Classifier.} The LOGAN \cite{Hayes2017LOGANMI} attack was originally a white box attack that was modified in \cite{vanbreugel2023membership} to a black box style and creates a surrogate model to approximate the target's characteristics by training a Generative Adversarial Network (GAN) using synthetic records. The discriminator $D_\theta(x)$ learns to distinguish between target-generated samples $S$ and reference dataset samples $R$, capturing the target model's distributional biases. For membership inference, the attack uses the learned discriminator function $f_{\text{LOGAN}}(x^*) = D_\theta(x^*)$ for each query record $x^*$, with the idea that member records should have a high probability of being assigned to the synthetic class. \cite{houssiau2022tapas} improves on this idea by instead training a supervised learning classifier such as a Random Forest rather than a GAN discriminator.

\textbf{Monte Carlo (MC).} The Monte Carlo attack \cite{Hilprecht2019MonteCA} exploits generative model overfitting by analyzing the density of generated samples around target records. This approach operates under the assumption that overfit generative models produce disproportionately more samples in areas surrounding their training data. The attack defines an $\varepsilon$-neighborhood around each query record $x^*$ as $U_\varepsilon(x^*) = {x' \mid d(x^*, x') \leq \varepsilon}$ and approximates the probability $P(s \in U_\varepsilon(x^*))$ via Monte Carlo integration. By taking $n$ samples $s_1, \ldots, s_n$ from $S$, the method computes the scoring function: $f_{\text{MC}}(x^*) = \frac{1}{n} \sum_{i=1}^{n} \mathbb{I}{(s_i \in U_\varepsilon(x^*))}$. This counting-based approach tallies generated samples within the $\varepsilon$-neighborhood of $x^*$, classifying records with higher density scores as likely training members.

\section{Is There a Strictly Dominant Attack for Synthetic Tabular Data?}
Given these attacks, a challenge facing adversaries attacking synthetic tabular data lies in selecting an MIA without prior knowledge of the underlying generative model or training data. Here, we hypothesize that different architectures, model initializations and training datasets can exhibit more or less of a specific privacy leakage signal which could influence with attacks see better performance. Given this unknown state for an adversary, we formally define the attack selection problem before running a massive experiment to answer Research Question 1.

\subsection{A Decision Theory Perspective on MIA Strategy Selection}
Given a No-box threat model, a synthetic dataset of unknown generator origin, and a reference dataset, the adversary must select a strategy with the goal of maximizing the discovered privacy leakage of a data publisher. 

\paragraph{Formal Setup} We model this as a decision problem under uncertainty where the state space $\Omega = (\mathcal{G}_1, \phi_1,  \mathcal{T}_1), (\mathcal{G}_2, \phi_2,  \mathcal{T}_2), \ldots,$ $ (\mathcal{G}_m, \phi_m, \mathcal{T}_n)$ represents all possible (generative model, parameter initialization, training dataset) combinations, the action space $\mathcal{A} = {A_1, A_2, \ldots, A_k}$ contains the available MIA strategies, and the payoff function $u: \mathcal{A} \times \Omega \rightarrow \mathbb{R}$ maps each (attack, state) pair to a performance measure (e.g., AUC, TPR at fixed FPR). The data publisher first commits to a state $\omega^* \in \Omega$ by selecting a generative model $\mathcal{G}$, initialization $\phi$,  and training dataset $\mathcal{T}$. The adversary, who only observes the synthetic data output and reference dataset under the No-Box threat model, must then choose an attack strategy $A_i \in \mathcal{A}$ without knowing the true state $\omega^*$.

While the space of possible (generative model, initialization, training dataset) combinations $\Omega$ is finite and observable ex-post through benchmark evaluation, the adversary must commit to an attack strategy ex-ante without knowledge of which specific combination they will encounter. This uncertainty creates a classic decision theory problem: how should a rational adversary choose among available attack strategies when the "state of the world" (i.e., the specific generative model, initialization and dataset) is unknown but the performance of each strategy under each possible state can be empirically evaluated post-hoc?

 Under this formulation, Research Question 1 asks whether there exists a strictly dominant strategy: $\exists A^* \in \mathcal{A}$ such that $u(A^*, \omega) \geq u(A_i, \omega) ; \forall A_i \in \mathcal{A}, \forall \omega \in \Omega$. If such a strategy exists, the adversary should always choose this attack as it would guarantee maximum success. Although it is theoretically difficult to show that any attack satisfies this condition, we conduct a massive preliminary experiment to test whether there is \textbf{ not} such a strategy.

 \begin{figure}
    \centering
    \includegraphics[width=\linewidth]{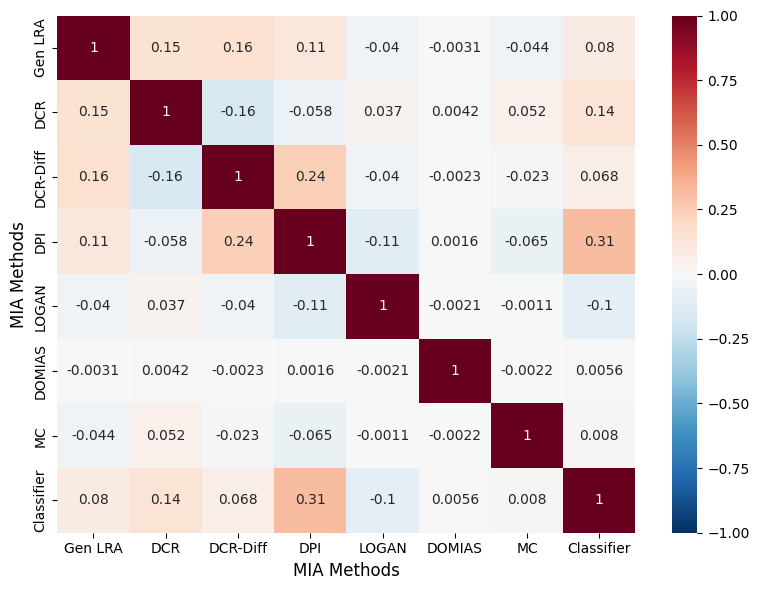}
    \caption{Mean correlations of various MIAs across datasets and seeds with synthetic data generated by TabSyn. While the scores of some MIAs are slightly correlated which each other, there is an overall diversity where different MIAs use different sources of signal in their methodology and thus see weak or no correlation with other strategies.}
    \label{fig:corr_plot}
    \Description{A correlation matrix heatmap showing the relationships between different membership inference attack (MIA) methods. The matrix displays correlation coefficients with values ranging from weak to moderate correlations, illustrating that different MIA strategies utilize distinct methodological approaches and signals.}

\end{figure}

\subsection{Experimental Design}\label{subsec: experiment design}
To empirically evaluate whether a strictly dominant MIA strategy exists, we construct the largest tabular synthetic data privacy experiment to date that spans a state space $\Omega$ of (generative model, dataset, seed) combinations. Our experimental design allows us to compute the payoff function $u(A_i, \omega)$ for each attack strategy $A_i$ across all observable states $\omega \in \Omega$, enabling us to test if a strategy $A^*$ does not achieve $u(A^*, \omega) \geq u(A_i, \omega)$ universally.

\paragraph{State Space Construction} 
We construct our state space $\Omega$ by combining 9 tabular generative models with 57 datasets across 5 seeds taken from a broad variety of fields including economics, healthcare, and social sciences, yielding 2565 distinct (model, seed, dataset) states. The generative models $\mathcal{G}$ include: CT-GAN, TVAE \cite{Xu2019ModelingTD}, Normalizing Flows (N-Flows) \cite{durkan2019neural}, Adversarial Random Forests (ARF) \cite{pmlr-v206-watson23a}, Tab-DDPM \cite{tabddpm}, PATEGAN \cite{yoon2018pategan}, AdsGAN \cite{yoon2020anonymization}, Auto-Diff \cite{suh2023autodiffcombiningautoencoderdiffusion}, and TabSyn \cite{tabsyn}. Our datasets $\mathcal{T}$ span 57 tabular datasets from the OpenML-CC18 Curated Classification benchmark \cite{oml-benchmarking-suites}, encompassing diverse domains and structural characteristics. As the original benchmark contains 72 datasets, we filter out instances that have greater than 100 columns as not all models can successfully handle such high dimensionality. All model implementations use default hyperparameters from Synthcity \cite{synthcity}, except Auto-Diff and TabSyn which use original codebases.
\begin{figure}
    \centering
    \includegraphics[width=1\linewidth]{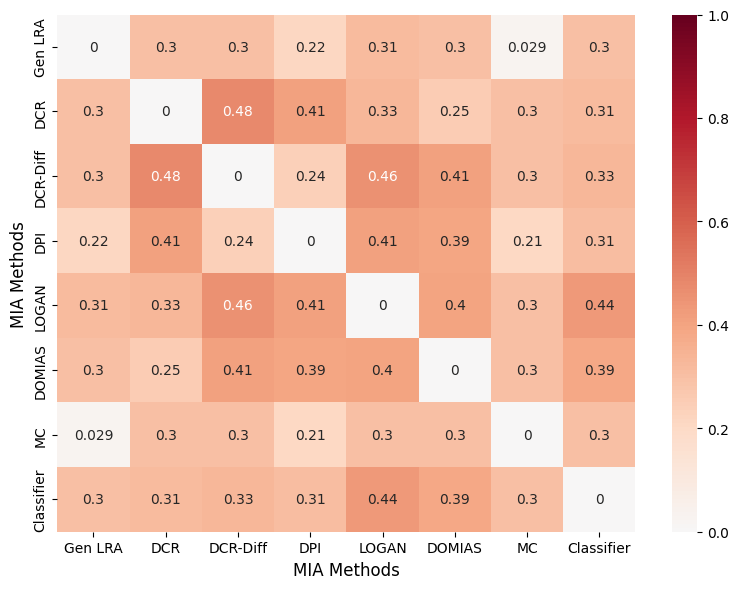}
    \caption{Mean disagreement rate of various MIAs across datasets and seeds with synthetic data generated by TabSyn. We threshold each attack by their median value, a heuristic used in \cite{ganleaks,vanbreugel2023membership}, and compare corresponding decisions. We find that many attacks often disagree with each other on between 20-40\% of observations.}
    \label{fig:disagreement}
    \Description{A matrix heatmap displaying pairwise disagreement rates between different membership inference attack methods. The matrix shows symmetric disagreement percentages, with most cells colored to indicate disagreement rates ranging from approximately 20-40\%, represented by varying color intensities across the grid of MIA method comparisons.}

\end{figure}
\begin{table*}[t]
\small
\centering
\caption{Rank comparison of individual attacks and ensembles. For each synthetic dataset we report the mean rank, top 3 proportion, and best proportion for each strategy. We find that ensemble strategies broadly see lower mean ranks for most success metrics.
}
\label{tab:performance_comparison}
\begin{tabular}{llccccccccc}
\toprule
\multirow{2}{*}{\textbf{Method}} & \multirow{2}{*}{\textbf{Type}} & \multicolumn{3}{c}{\textbf{AUC}} & \multicolumn{3}{c}{\textbf{TPR@0.01}} & \multicolumn{3}{c}{\textbf{TPR@0.1}} \\
\cmidrule(lr){3-5} \cmidrule(lr){6-8} \cmidrule(lr){9-11}
& & \textbf{MeanRank $\downarrow$} & \textbf{PTop3 $\uparrow$} & \textbf{PBest $\uparrow$} & \textbf{MeanRank $\downarrow$} & \textbf{PTop3 $\uparrow$} & \textbf{PBest $\uparrow$} & \textbf{MeanRank $\downarrow$} & \textbf{PTop3 $\uparrow$} & \textbf{PBest $\uparrow$} \\
\midrule
Weighted Mean & Ensemble & \textbf{3.683 (0.437)} & \textbf{0.447} & 0.066 & 4.166 (0.480) & 0.413 & 0.118 & 4.254 (0.388) & 0.420 & 0.108 \\
Majority Voting & Ensemble & 3.706 (0.423) & 0.436 & 0.094 & 4.284 (0.423) & \textbf{0.443} & 0.030 & \textbf{4.085 (0.401)} & \textbf{0.452} & 0.082 \\
Mean & Ensemble & 4.303 (0.462) & 0.378 & 0.061 & \textbf{4.091 (0.485)} & 0.441 & 0.086 & 4.363 (0.368) & 0.472 & 0.066 \\

DPI & Individual & 5.455 (0.424) & 0.314 & 0.108 & 4.918 (0.403) & 0.378 & \textbf{0.136} & 5.189 (0.430) & 0.351 &\textbf{ 0.138} \\
DCR & Individual & 5.668 (0.452) & 0.317 &\textbf{ 0.120} & 5.595 (0.424) & 0.299 & 0.107 & 5.574 (0.442) & 0.315 & 0.100 \\
DOMIAS & Individual & 5.990 (0.414) & 0.241 & 0.093 & 5.377 (0.437) & 0.345 & 0.118 & 5.766 (0.428) & 0.287 & 0.080 \\
Classifier & Individual & 6.155 (0.470) & 0.294 & 0.114 & 5.726 (0.423) & 0.292 & 0.090 & 5.939 (0.459) & 0.293 & 0.106 \\
Gen-LRA & Individual & 6.484 (0.466) & 0.246 & 0.109 & 6.093 (0.422) & 0.238 & 0.086 & 6.241 (0.445) & 0.245 & 0.093 \\
MC & Individual & 6.722 (0.457) & 0.234 & 0.061 & 6.037 (0.432) & 0.261 & 0.086 & 6.438 (0.461) & 0.244 & 0.093 \\
LOGAN & Individual & 6.864 (0.439) & 0.191 & 0.066 & 6.537 (0.326) & 0.108 & 0.014 & 6.621 (0.386) & 0.168 & 0.031 \\
DCR-Diff & Individual & 7.837 (0.518) & 0.208 & 0.111 & 6.431 (0.463) & 0.245 & 0.101 & 7.130 (0.488) & 0.217 & 0.106 \\
\bottomrule
\end{tabular}
\end{table*}
\paragraph{Data Generation} Following standard synthetic data benchmarking practices \cite{synthcity,tabsyn}, the dataset is split into 80:20  train/test partitions, and the tabular synthetic data generator is fit to the training partition. A synthetic dataset is then generated to match the original size of the training dataset. To account for randomness in model training and sampling, each experimental configuration is repeated across five independent runs. Following the recommendations of prior work \cite{guépin2024lostaveragesnewspecific}, we fix the train/test partition across all runs and vary only the generative model initialization seeds. This design helps isolate the variability due to model behavior from that due to evaluation set construction, which is especially important in privacy attack scenarios.

\paragraph{MIA Setup} To evaluate each MIA, we further split the test partition into equal size holdout and reference sets. All data is then encoded based on the synthetic dataset to prevent data leakage. We scale continuous variables, one-hot encode categorical variables for distance-based attacks, and ordinally encode them for KDE-based attacks. Each MIA then evaluates a test dataset which is the union of the training and holdout partitions using the available reference and synthetic sets as prescribed by the threat model. 

\paragraph{MIA Evaluation} Throughout this paper, we evaluate MIAs based on their relative rank performance over many different states. This is in contrast to MIA evaluation procedures that often use just a handful of datasets and compare the performance of the methods conditioned on each dataset. While aggregating success by means can under-report extreme success or failure in individual states \cite{guépin2024lostaveragesnewspecific}, relative rank over multiple states provides a more robust assessment of method performance. This approach allows us to capture the consistency of MIA effectiveness across diverse conditions and reduces the risk of drawing conclusions based on dataset-specific artifacts or outliers. By examining relative rankings rather than absolute performance metrics, we can better understand which methods demonstrate superior performance across the full spectrum of evaluation scenarios, leading to more generalizable insights about MIA capabilities. We primarily evaluate rankings for AUC and TPR at low fixed FPR, which has become standard as a measure of the "meaningful effectiveness" of an attack \cite{Carlini2021MembershipIA}.

\subsection{Results}

For each state across all datasets, models, and random seeds, we plot the proportion of states each attack has the highest effectiveness in Figure \ref{fig:att_prop}. Overall, we find the distribution of the rank 1 attack success is remarkably uniform with the best attack, DPI, only seeing the top AUC and TPR@FPR=.01 ranks in 16.2\% and 19.1\% of states respectively. While this performance is impressive for DPI, it implies that in the large majority of states, if an adversary used DPI they would not have achieved the empirically best attack performance possible.  

To measure the similarity of these different strategies, we plot the pairwise correlation and disagreement of attacks over an example state of the Credit dataset generated by TabSyn in Figures \ref{fig:corr_plot} and \ref{fig:disagreement}. We find that the pairwise attack sample-level scores are often weakly correlated and the classification decisions of these attacks have high disagreement. This implies that different attack strategies are indeed targeting different signal sources of privacy leakage.

These findings have important implications for privacy risk assessment in synthetic data generation. The relatively uniform distribution of maximal attack effectiveness across states and the weak correlations between different attack strategies suggest that no single attack provides a strictly dominant evaluation of privacy leakage. This means that relying on any one attack strategy may systematically underreport the actual privacy risks present in a synthetic dataset, as each method appears to exploit different vulnerabilities in the data generation process. Furthermore, the high disagreement between attack classifications introduces an additional layer of complexity: the privacy risk for any individual sample becomes contingent on which attack strategy an adversary might choose to employ. This variability underscores the need for comprehensive privacy evaluation frameworks that incorporate multiple attack vectors rather than relying on single-method assessments, as the true privacy landscape can only be understood through the lens of diverse adversarial approaches.

\section{Ensembling MIAs}
In the absence of a strictly dominant attack, a natural question arises: what course of action should a rational adversary take? While an adversary could naively default to a single method like DPI, the empirical diversity we observed across individual attacks suggests a more sophisticated approach may be warranted. This leads us to Research Question 2: Can ensemble methods create more robust MIA strategies that minimize regret compared to individual attacks? The lack of a universally optimal strategy, combined with the complementary strengths exhibited by different individual attacks, motivates us to explore whether these methods can be treated as weak learners and combined through unsupervised ensembling techniques. In this section, we investigate how ensemble approaches can leverage the diverse signals from individual attacks to provide more consistent and robust performance across varying states.
\begin{figure*}
    \centering
    \includegraphics[width=1\linewidth]{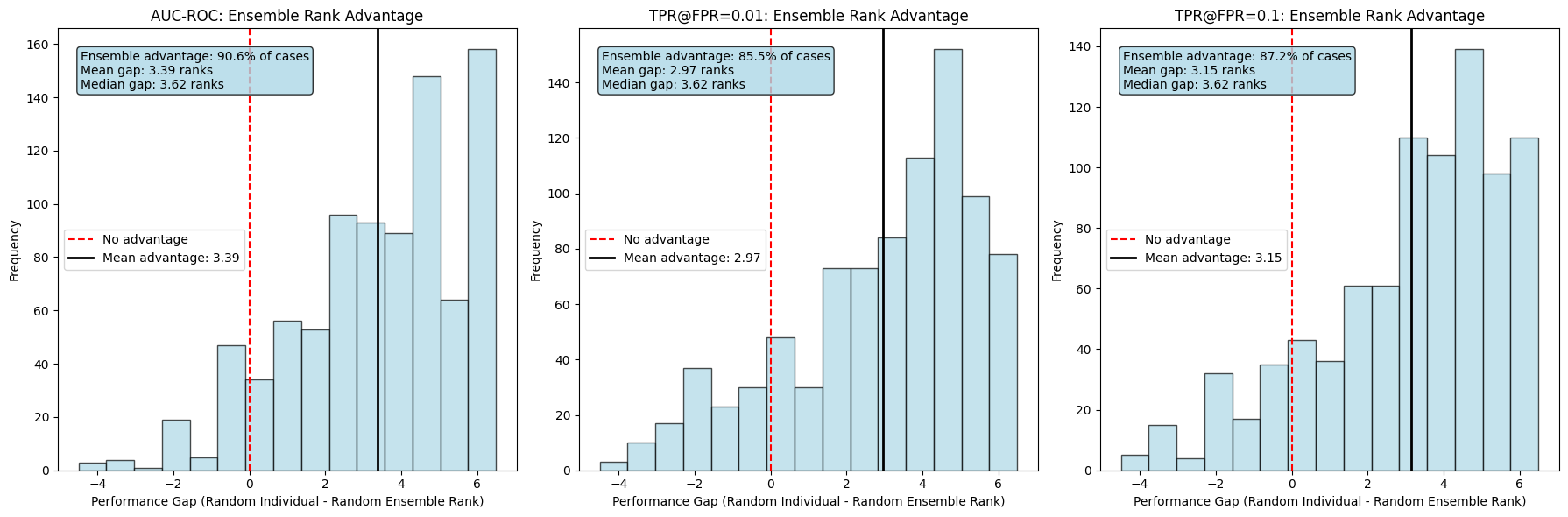}
    \caption{Advantage distributions for various success metrics. In the absence of a dominant strategy, we compare a random ensemble to a random individual attack and plot the improvement in overall rank for if an adversary had selected that ensemble vs that attack. We find that selecting an ensemble improves the rank of an adversary's strategy an average of 3.15 ranks when evaluated over TPR@FPR=0.1.}
    \label{fig:advantage}
        \Description{Multiple box plots arranged horizontally showing the distribution of rank advantage improvements across different success metrics. Each plot displays the statistical distribution of rank improvements when comparing ensemble attacks to individual attacks, with the plots showing varying spread and central tendencies for different evaluation metrics.}

\end{figure*}
\subsection{MIAs as Weak Learners}

In machine learning, weak learners are classifiers that perform only marginally better than random guessing, yet still possess predictive signal. Formally, a weak learner $h: \mathcal{X} \rightarrow \{0,1\}$ satisfies $\mathbb{P}[h(x) = y] \geq \frac{1}{2} + \gamma$ for some advantage $\gamma > 0$, where $\gamma$ represents the margin above random performance. This is typically characterized by marginal accuracy improvements, high variance across different conditions, and limited individual discriminative power \cite{Schapire2004The, Kolter2023Language, George2006Bayesian, Lee2009A}. 

Weak learners serve as fundamental building blocks that can be combined to create strong learners through ensemble methods that exploit their diversity. Given a set of weak learners $\{h_1, h_2, \ldots, h_T\}$, an ensemble method produces a strong learner: \[
H(x) = \operatorname{sign}\!\left(\sum_{t=1}^T \alpha_t h_t(x)\right)
\] where $\alpha_t \geq 0$ are the weights assigned to each weak learner. The effectiveness of ensemble methods fundamentally depends on diversity among base learners, quantified by $\mathbb{E}_{x \sim \mathcal{D}}[\mathbb{I}[h_i(x) \neq h_j(x)]] > 0$ for $i \neq j$, which enables uncorrelated individual errors to cancel out through aggregation mechanisms \cite{Zhou2021An, Liu2011Create}. 
Theoretical analyses demonstrate that ensemble error decreases as correlation between individual learner errors decreases \cite{Yao1999Ensemble, Medarhri2025Constructing, Cheung2025Error}, making individual MIAs well-suited for ensemble effectiveness. 

\begin{table*}[ht]
\centering
\caption{Mean rank contribution (with standard error) of each individual attack across all states and ensembles. For each MIA, we compute While DPI sees good individual performance in previous experiments, we find that "weaker" attacks see higher contribution to the success of ensembles. This indicates that these individual attacks are useful in their ability to construct better ensembles.}

\begin{tabular}{lccc
}
\toprule
\textbf{Method} & \textbf{AUC Contr. $\downarrow$}  & \textbf{TPR@FPR.01 Contr. $\downarrow$} & \textbf{TPR@FPR.1 Contr. $\downarrow$} \\
\midrule
DCR         & \textbf{2.66 (0.42)} & \textbf{3.34 (0.51)} & 4.14 (0.51) \\
DCR-Diff    & 3.79 (0.32) & 3.52 (0.45) & \textbf{3.24 (0.34)} \\
Gen-LRA    & 4.21 (0.42) & 4.00 (0.56) & 3.83 (0.42) \\
MC             & 4.31 (0.39) & 4.10 (0.43) & 4.45 (0.43) \\
DPI     & 4.76 (0.50) & 4.07 (0.35) & 5.21 (0.41) \\
DOMIAS         & 5.21 (0.22) & 3.93 (0.42) & 4.76 (0.29) \\
LOGAN          & 5.38 (0.32) & 3.90 (0.38) & 4.59 (0.37) \\
Classifier     & 5.66 (0.50) & 5.28 (0.45) & 5.00 (0.53) \\
\bottomrule
\end{tabular}
\label{tab:contribution}
\end{table*}

\subsection{Unsupervised Ensembles}
Having established that individual MIA can function as weak learners with complementary strengths and diverse error patterns, we now examine three unsupervised ensemble methods that can aggregate their predictions without requiring additional training data. These methods directly exploit the diversity properties identified above to create more robust inference strategies.

Consider a collection of $N$ individual MIA strategies $\{A_1, A_2, \ldots,$ $A_N\}$, where each attack $A_a$ produces a membership inference score $s_{ia} \in \mathbb{R}$ for data point $i$. The score vector $\mathbf{s}_i = [s_{i1}, s_{i2}, \ldots, s_{iN}]$ represents the aggregated output from all $N$ attacks on point $i$, where higher scores typically indicate stronger evidence of membership. We explore several ensemble methods combine these individual attack scores to produce a final inference decision that leverages the collective intelligence of the diverse MIA strategies.

\textbf{Mean Ensemble} aggregates attack scores through simple arithmetic averaging, treating all attackers equally in the final prediction. The ensemble score is computed as:
$$\text{Mean}(i) = \frac{1}{N}\sum_{a=1}^{N} s_{ia}$$
This approach assumes that all attackers provide equally reliable predictions and that errors are randomly distributed across attackers, allowing them to cancel out through averaging. While computationally efficient and interpretable, mean ensemble can be sensitive to outlier scores from poorly calibrated attackers, as extreme values directly influence the final prediction without any normalization or weighting mechanism.

\textbf{Weighted Mean Ensemble} extends the basic mean approach by incorporating attacker-specific weights that reflect their individual performance or reliability. The ensemble score incorporates predetermined weights $w_a$ for each attacker $a$:
$$\text{WeightedMean}(i) = \frac{\sum_{a=1}^{N} w_a \cdot s_{ia}}{\sum_{a=1}^{N} w_a}$$
where weights $w_a$ are typically derived from validation performance metrics such as AUC, accuracy, or precision-recall measures. This formulation allows high-performing attackers to contribute more significantly to the final prediction while maintaining contributions from all ensemble members if prior information is known. Here, we assign a weight vector based on Figure \ref{fig:att_prop} where each attack is given a weighting based on the proportion of states that attack achieved the best AUC as given these experiments are public and adversary could now use this information as some set of priors.

\textbf{Majority Voting Ensemble} converts continuous attack scores into binary membership predictions and aggregates them through democratic voting. Each attacker $a$ first converts its score $s_{ia}$ into a binary decision $b_{ia}$ using a threshold $\tau_a$:
$$b_{ia} = \begin{cases} 1 & \text{if } s_{ia} \geq \tau_a \\ 0 & \text{otherwise} \end{cases}$$
The final ensemble prediction is determined by majority consensus:
$$\text{MajorityVote}(i) = \begin{cases} 1 & \text{if } \sum_{a=1}^{N} b_{ia} > \frac{N}{2} \\ 0 & \text{otherwise} \end{cases}$$
This approach transforms the membership inference problem into a discrete voting scenario where each attacker contributes an equal vote. In our experimentation, we threshold the scores based on the median value. Majority voting is robust to individual attacker failures and provides interpretable results, but requires careful threshold selection for each attacker to ensure balanced voting behavior across the ensemble.

\section{Ensemble Performance}
To evaluate the performance of ensembled MIAs for tabular generative models, we repeat the experiment from Section \ref{subsec: experiment design}, but now include each introduced method. For each ensemble, we use as input one of each attack and compare the relative performance of each ensemble and individual attack for each state.
\subsection{Ensemble Success}

We primarily evaluate the ensembles using a relative rank-based methodology which provides several advantages for ensemble evaluation. First, it treats each synthetic dataset as an independent evaluation scenario, giving equal weight to performance across different datasets and generation methods. Second, it directly answers the practical question: "Given an arbitrary synthetic dataset of unknown provenance, which attack strategy is most likely to yield near-optimal results?" Finally, by focusing on relative rather than absolute performance differences, this approach remains robust to variations in dataset difficulty and inherent privacy vulnerabilities across different tabular domains.

We report the Mean Relative Rank with standard error and the proportion of synthetic datasets where each method ranked in the top 3 (PTop3) and achieved the best performance (PBest) across AUC, TPR@FPR=0.01, and TPR@FPR=0.1 metrics in Table \ref{tab:performance_comparison}. Overall, ensembles demonstrate improved performance over individual attacks in terms of mean rank and PTop3 across all metrics. However, no ensemble achieves a higher PBest than individual attacks. This indicates that while ensembles perform more consistently, some individual attacks still achieve the highest relative empirical performance across more states.

Although unsupervised ensembles are not always optimal, they effectively leverage diverse signals to provide greater average advantage for an adversary. The superior mean rank and PTop3 performance of ensembles directly translates to minimized regret in practical scenarios. Since an adversary cannot know a priori which individual attack will perform best on a given dataset, selecting an individual attack risks poor performance when that specific method fails. In contrast, ensembles' consistently higher PTop3 scores demonstrate their ability to maintain competitive performance across diverse conditions, while their improved mean ranks show they avoid the worst-case scenarios that individual attacks may encounter. Therefore, across all experimental conditions, an adversary would minimize their expected regret by selecting an ensemble approach, trading the possibility of achieving the absolute best performance for the guarantee of consistently strong results regardless of dataset characteristics.

We further evaluate this additional advantage for rank performance in Figure \ref{fig:advantage}. For each run we compare the difference in rank for AUC, TPR@FPR=0.01, TPR@FPR=0.1 between a randomly selected individual attack and random ensemble. We find that for 90.6\% of synthetic datasets a random ensemble outperforms the AUC of a random individual attack and sees a mean rank improvement of 3.39. This demonstrates that for an adversary without strong priors for which individual method will perform best, ensembling will usually improve their attack.
\begin{figure*}[t]
    \centering
    \begin{subfigure}{0.45\textwidth}
        \centering
        \includegraphics[width=\linewidth]{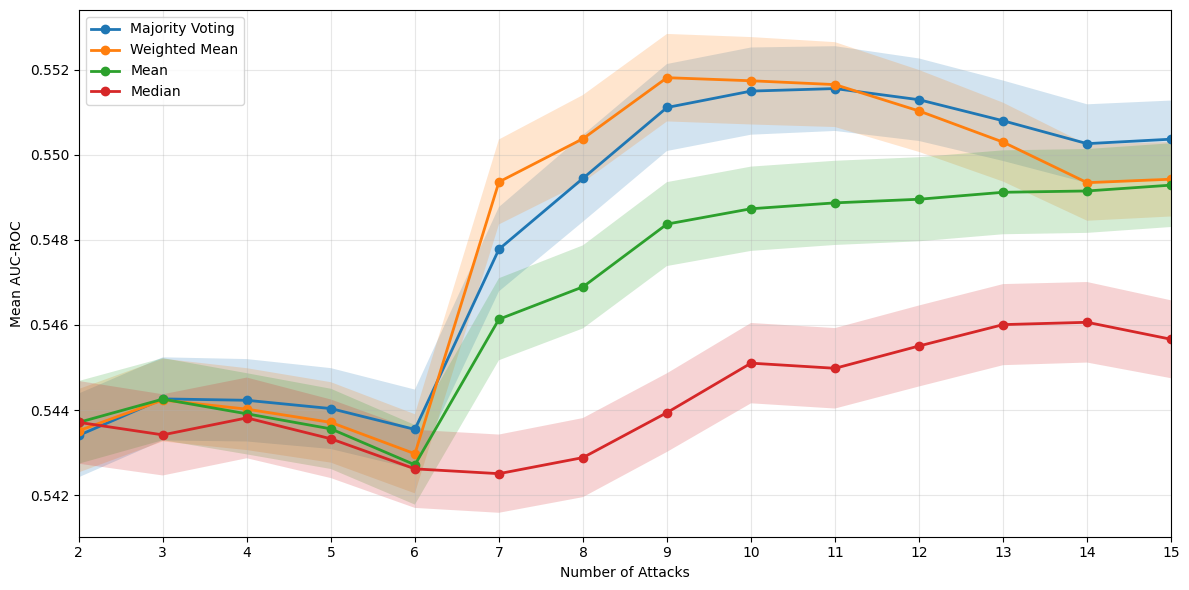}
        \label{fig:auc_num_att}

    \end{subfigure}
    \hfill
    \begin{subfigure}{0.45\textwidth}
        \centering
        \includegraphics[width=\linewidth]{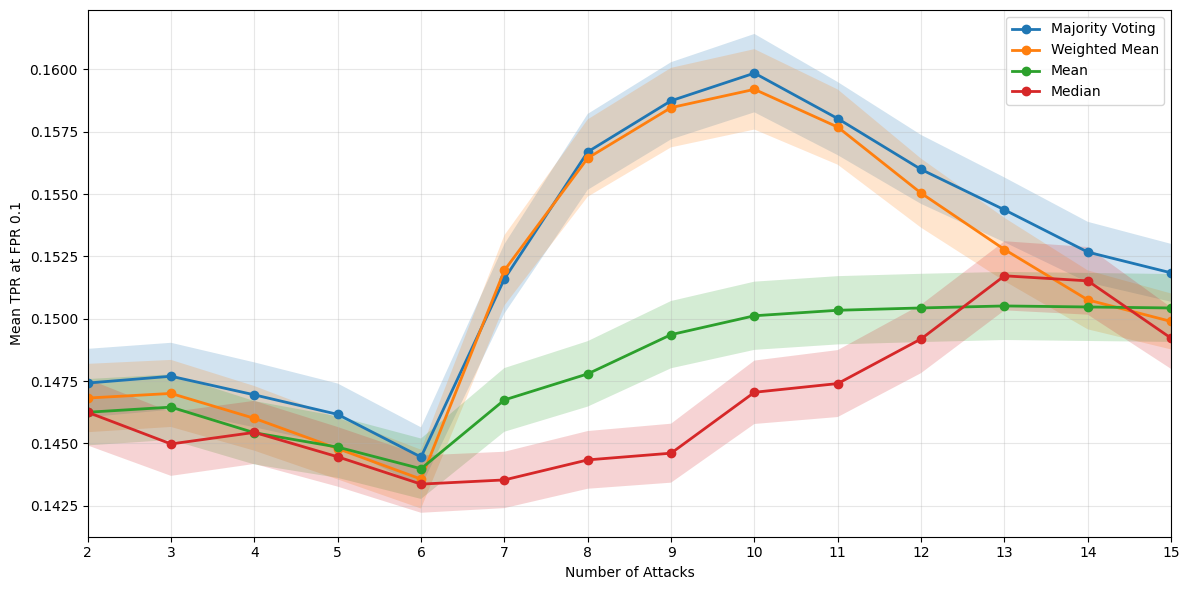}
        \label{fig:tpr_at_fpr_num_att}

    \end{subfigure}
    \caption{Mean AUC and TPR@FPR=0.1 for Majority Voting as ensemble size grows. We find that ensembles see improvement at around 7 attacks and see diminishing returns after 10 attacks. This is likely because as attacks become repeated with different hyperparameter settings there is little new signal for the ensemble to exploit.}
    \Description{Two side-by-side line graphs showing performance metrics versus ensemble size with multiple lines representing each ensemble. The left graph displays AUC values and the right graph shows TPR@FPR=0.1 values, both plotted against the number of attacks in the ensemble. Both graphs show curves that rise initially and then level off, demonstrating the relationship between ensemble size and performance metrics.}

    \label{fig:combined_performance}
\end{figure*}

\subsection{Attack Contribution}

Under ensembling, the value of a strategy for an adversary is not solely determined by its individual performance across states, but rather by its marginal contribution to ensemble performance. We employ a leave-one-out analysis scheme to quantify each attack's contribution to ensemble performance across all evaluated states. Our methodology proceeds as follows: For an ensemble containing $n$ attacks, we construct $n$ reduced ensembles, each excluding exactly one constituent attack. For each state, we compute the performance for both the complete ensemble and each reduced variant. The marginal contribution of attack is defined as the difference between the complete ensemble's success metric and the success metric of the ensemble excluding attack.

Formally, if $E$ represents the complete ensemble and $E_{-a}$ represents the ensemble excluding attack $a$, then the marginal contribution $C_{a,s}$ for attack $a$ in state $\omega$ over an evaluation function $u(\cdot)$ is:

\begin{equation}
C_{a,\omega} = u(E,\omega) - u(E_{-a}, \omega)
\end{equation}

We report the mean rank contribution of each individual attack for all states in Table \ref{tab:contribution}. Here, for each ensemble run, we compute the leave-one-out contribution of each MIA by measuring the change in ensemble performance when that attack is excluded. We then rank these contributions and report the mean rank across ensemble types and runs. We find that overall, attacks that did not excel individually, such as DCR, DCR-Diff, and Gen-LRA contributed relatively more on average to the performance of the ensemble on AUC and TPR at Fixed FPR than the best individual attack DPI. This demonstrates that sub-optimal individual strategies can be useful privacy auditing so long as they are sufficiently uncorrelated to improve the performance of the ensemble.

\subsection{Including Additional Attacks} \label{subsec: Additional Attacks}
Ensembles can incorporate any number of individual attacks as input components. To understand how ensemble performance scales with the diversity and quantity of constituent attacks, we systematically evaluate ensembles of varying sizes by randomly selecting a growing number of individual attacks with different hyperparameter initializations.

Our experimental design samples attack combinations ranging from 2 attacks to larger collections of up to 25, with each attack using different hyperparameter configurations to maximize diversity in the ensemble's constituent strategies. This approach allows us to investigate two key questions: whether additional attacks consistently improve ensemble performance, and at what point diminishing returns become apparent. We repeat this MIA randomization for 100 runs and report the mean AUCs and TPR@FPR=.1 for all synthetic datasets and report the performance of various ensembles in Figure \ref{fig:combined_performance}.

We find that for both AUC and TPR@FPR=0.1, ensemble strategies see improvements after 7 or more individual attacks are included and see gains until approximately 11 attacks. As we add more attacks, attacks get repeated but with different instantiations of hyperparameters which likely begin to not contribute additional signal to the ensemble due to their correlation with same attack at different hyperparameters. An additional advantage of ensembles is that any new attack created in the future can improve these methods provided that it is approximately orthogonal to existing attacks, i.e. it increases the diversity of the ensemble.

\section{Discussion}
\subsection{Practical Privacy Implications}
Our systematic evaluation reveals that no single membership inference attack consistently dominates across all generative models and datasets, creating a complex landscape of privacy vulnerabilities in synthetic data generation. While individual generative models may exhibit resistance or vulnerability to specific MIAs, our results demonstrate that ensemble-based attack strategies achieve superior long-term performance across diverse experimental conditions compared to any individual attack method.

These findings carry several critical implications for privacy auditing and defense strategies in synthetic data systems. First, practitioners conducting privacy evaluations should deploy comprehensive attack portfolios rather than relying on single-method assessments when seeking to quantify maximum empirical privacy leakage. Our results show that any individual strategy is empirically unlikely to represent the worst-case scenario a defender might encounter in their specific deployment context. This principle extends to ensemble methods themselves, as each ensemble configuration achieved optimal performance in 5-10\% of experimental states, underscoring the importance of an auditor deploying many evaluation approaches.

Second, defensive strategies and evaluation frameworks— including similarity-based metrics—that focus exclusively on mitigating individual attack types prove insufficient in practice. The superior effectiveness of ensemble methods indicates that adversaries can exploit multiple, potentially orthogonal vulnerability signals to circumvent defenses optimized against specific attack patterns. This has profound implications for privacy-preserving synthetic data generation: robust defenses must account for the complete attack surface rather than optimizing against isolated methods. While our study focuses on popular non-differentially private synthetic data generators, these findings highlight the significant potential value of differential privacy \cite{dwork2006calibrating} as a comprehensive defense mechanism.

Finally, our results suggest that ensemble attacks may represent a more realistic and immediate threat model for data publishers than theoretically optimal individual attacks. Since adversaries cannot determine a priori which attack will perform optimally on a given dataset, ensemble strategies offer a more practical and achievable threat vector. This paradigm shift—from defending against hypothetically perfect attacks to mitigating consistently strong ensemble approaches—provides a more actionable framework for privacy risk assessment and mitigation in real-world synthetic data deployment scenarios.

\subsection{Prioritize Signal Diversity for Future Attacks}

The effectiveness of ensemble approaches fundamentally shifts the evaluation paradigm for novel membership inference attacks, creating new opportunities for attack development that transcend traditional performance-centric metrics. Rather than requiring new attacks to achieve state-of-the-art individual performance, ensemble frameworks value attacks that contribute unique privacy leakage signals, even when their standalone performance remains modest. When these diverse signals exhibit weak or no correlation, they provide complementary information that can substantially enhances overall ensemble effectiveness.

This perspective carries important implications for the privacy research community. Researchers can focus on developing attacks that target previously unexplored privacy leakage mechanisms without the traditional constraint of achieving competitive standalone performance. An attack that meaningfully improves an already competitive ensemble strategy represents a valuable contribution to the adversarial toolkit, regardless of its individual performance metrics. This framework encourages exploration of novel vulnerability surfaces and attack vectors that might otherwise be overlooked in individual performance-focused evaluation paradigms.

\section{Conclusion}
This work introduces a fundamental challenge in privacy auditing for tabular synthetic data: the absence of a universally effective membership inference attack. Through the largest systematic evaluation of MIA performance to date, spanning 9 generative models and 57 datasets, we demonstrate that no single attack consistently dominates across diverse experimental conditions and a realistic threat model.

Our framing of synthetic data MIAs as a decision-theoretic problem under uncertainty reveals that ensemble-based MIA strategies offer superior regret-minimizing performance compared to individual attacks. These ensemble approaches consistently achieve better mean ranks across our comprehensive benchmark, providing more robust privacy assessment tools for practitioners. Importantly, we show that even attacks with modest standalone performance can contribute significantly to ensemble effectiveness.

This work opens promising directions for future research. First, the development of more sophisticated ensemble architectures presents opportunities to improve upon the unsupervised methods demonstrated here, potentially incorporating adaptive weighting schemes or hierarchical attack combinations. Second, the value of signal diversity motivates systematic exploration of uncorrelated individual MIAs that target previously unexplored privacy leakage mechanisms, as even modestly performing attacks can enhance ensemble effectiveness. These research directions can lead to more comprehensive privacy auditing methodologies.

\begin{acks}
This work was supported in part by the National Science Foundation under Grant CNS-2247795 and by the Office of Naval Research under Grant N00014-22-1-2680. Any opinions, findings, and conclusions or recommendations expressed in this material are those of the authors and do not necessarily reflect the views of the National Science Foundation or the Office of Naval Research.
\end{acks}
\bibliographystyle{ACM-Reference-Format}
\bibliography{sample-base}

\section{Appendix}
\subsection{Metric Definitions}
\subsubsection{AUC-ROC (Area Under the Receiver Operating Characteristic Curve)}
The area under the curve formed by plotting the True Positive Rate (TPR) against the False Positive Rate (FPR) at various classification thresholds. Mathematically:
$$\text{AUC-ROC} = \int_0^1 \text{TPR}(\text{FPR}^{-1}(x)) \, dx$$
where TPR = TP/(TP+FN) and FPR = FP/(FP+TN). Values range from 0 to 1, with 0.5 indicating random performance and 1.0 indicating perfect classification.

\subsubsection{TPR@Fixed FPR (True Positive Rate at Fixed False Positive Rate)}
The true positive rate achieved when the false positive rate is constrained to a specific value $\alpha$:
$$\text{TPR@FPR}_\alpha= \max_\theta \{\text{TPR}(\theta) : \text{FPR}(\theta) \leq \alpha\}$$
where $\theta$ represents the classification threshold. This metric is particularly useful when controlling for acceptable false positive rates in applications with asymmetric costs.

\subsubsection{Mean Rank}
For a ranking task with n items, the average position of relevant items in the ranked list:
$$\text{Mean Rank} = \frac{1}{|R|} \sum_{i \in R} \text{rank}(i)$$
where R is the set of relevant items and rank(i) is the position of item i in the ranked list (typically starting from 1). Lower values indicate better ranking performance.
\subsection{Datasets}
We report the data sets used for the experiments in Sections 3-5 in Table \ref{tab:datasets}.
\begin{table}[ht]
    \caption{List of OpenML datasets included in the experiments}
    \centering
    \tiny
\begin{tabular}{lrrrrr}
\hline
Dataset & OpenML ID & N-size & Classes &  Cat. Feat. &  Num Feat. \\
\hline
GesturePhaseSegmentationProcessed & 4538 & 9873 & 5 & 1 & 32 \\
MiceProtein & 40966 & 1080 & 8 & 5 & 77 \\
PhishingWebsites & 4534 & 11055 & 2 & 31 & 0 \\
adult & 1590 & 48842 & 2 & 9 & 6 \\
analcatdata\_authorship & 40983 & 4839 & 2 & 1 & 5 \\
analcatdata\_dmft & 469 & 797 & 6 & 5 & 0 \\
bank-marketing & 1461 & 45211 & 2 & 10 & 7 \\
banknote-authentication & 1462 & 1372 & 2 & 1 & 4 \\
blood-transfusion-service-center & 1464 & 748 & 2 & 1 & 4 \\
breast-w & 15 & 699 & 2 & 1 & 9 \\
car & 40975 & 1728 & 4 & 7 & 0 \\
churn & 40701 & 5000 & 2 & 5 & 16 \\
climate-model-simulation-crashes & 1467 & 540 & 2 & 1 & 20 \\
cmc & 23 & 1473 & 3 & 8 & 2 \\
connect-4 & 40668 & 67557 & 3 & 43 & 0 \\
credit-approval & 29 & 690 & 2 & 10 & 6 \\
credit-g & 31 & 1000 & 2 & 14 & 7 \\
cylinder-bands & 6332 & 540 & 2 & 22 & 18 \\
diabetes & 37 & 768 & 2 & 1 & 8 \\
dresses-sales & 23381 & 500 & 2 & 12 & 1 \\
electricity & 151 & 45312 & 2 & 2 & 7 \\
eucalyptus & 43924 & 736 & 5 & 15 & 5 \\
first-order-theorem-proving & 1475 & 6118 & 6 & 1 & 51 \\
ilpd & 1480 & 583 & 2 & 2 & 9 \\
jm1 & 1053 & 10885 & 2 & 1 & 21 \\
kc1 & 1067 & 2109 & 2 & 1 & 21 \\
kc2 & 1063 & 522 & 2 & 1 & 21 \\
kr-vs-kp & 3 & 3196 & 2 & 37 & 0 \\
letter & 6 & 20000 & 26 & 1 & 16 \\
mfeat-fourier & 14 & 2000 & 10 & 1 & 76 \\
mfeat-karhunen & 16 & 2000 & 10 & 1 & 64 \\
mfeat-morphological & 18 & 2000 & 10 & 1 & 6 \\
mfeat-zernike & 22 & 2000 & 10 & 1 & 47 \\
numerai28.6 & 23517 & 96320 & 2 & 1 & 21 \\
optdigits & 28 & 5620 & 10 & 1 & 64 \\
ozone-level-8hr & 1487 & 2534 & 2 & 1 & 72 \\
pc3 & 1044 & 10936 & 3 & 4 & 24 \\
pendigits & 32 & 10992 & 10 & 1 & 16 \\
phoneme & 1489 & 5404 & 2 & 1 & 5 \\
qsar-biodeg & 1494 & 1055 & 2 & 1 & 41 \\
satimage & 182 & 6430 & 6 & 1 & 36 \\
segment & 40984 & 2310 & 7 & 1 & 19 \\
sick & 38 & 3772 & 2 & 23 & 7 \\
spambase & 44 & 4601 & 2 & 1 & 57 \\
splice & 46 & 3190 & 3 & 62 & 0 \\
steel-plates-fault & 40983 & 4839 & 2 & 1 & 5 \\
texture & 40499 & 5500 & 11 & 1 & 40 \\
tic-tac-toe & 50 & 958 & 2 & 10 & 0 \\
vehicle & 54 & 846 & 4 & 1 & 18 \\
\hline
\end{tabular}
    \label{tab:datasets}
\end{table}

\subsection{Further Experiment Details for Section \ref{subsec: Additional Attacks}}

For Section \ref{subsec: Additional Attacks}, we report the hyperparameters for each possible instantiation of each attack. A random selection of $N$ attack + hyperparameter settings are taken from this list to be used for the ensemble and are processed in accordance with the details from Section \ref{subsec: experiment design}.

\begin{itemize}
    \item \textbf{DCR:} L1 and L2 distance
    \item \textbf{DCR-Diff:} L1 and L2 distance  
    \item \textbf{Gen-LRA:} $K \in \{1, 3, 5, 10, 20, 50\}$
    \item \textbf{DPI:} $K \in \{1, 3, 5, 10, 20, 50\}$
    \item \textbf{Classifier:} Model $\in \{\text{RandomForest}, \text{XGBoost}, \text{Log. Reg.}\}$
    \item \textbf{MC/LOGAN/DOMIAS:} default parameters
\end{itemize}
\end{document}